\documentclass[range]{ar2e}

\def\simgt{\lower 2pt \hbox{$\, \buildrel {\scriptstyle >}\over {\scriptstyle \sim}\,$}}
\def\simlt{\lower 2pt \hbox{$\, \buildrel {\scriptstyle <}\over {\scriptstyle \sim}\,$}}
\def\arielv{{\it Ariel~V\/}}
\def\asca{{\it ASCA\/}}
\def\chandra{{\it Chandra\/}}
\def\conx{{\it Constellation-X\/}}
\def\einstein{{\it Einstein\/}}
\def\genx{{\it Generation-X\/}}
\def\heao1{{\it {\it HEAO-1}\/}}
\def\hst{{\it {\it HST}\/}}

\def\rosat{{\it ROSAT\/}}
\def\rxte{{\it RXTE\/}}
\def\sax{{\it BeppoSAX\/}}
\def\spitzer{{\it Spitzer\/}}

\def\uhuru{{\it Uhuru\/}}
\def\wmap{{\it WMAP\/}}
\def\xeus{{\it XEUS\/}}
\def\xmm{{\it XMM-Newton\/}}

\begin{document}

\input epsf.tex 
\input epsf.def 
\input psfig.sty

\jname{Annual Reviews of Astronomy \& Astrophysics}
\jyear{2004}
\jvol{43}
\ARinfo{1056-8700/97/0610-00}

\title{Deep Extragalactic X-ray Surveys}

\markboth{Deep Extragalactic X-ray Surveys}{Deep Extragalactic X-ray Surveys}

\author{W.N. Brandt$^1$ and G. Hasinger$^2$
\affiliation{$^1$Department of Astronomy \& Astrophysics, The Pennsylvania
State University, 525 Davey Lab, University Park, Pennsylvania 16802, USA; 
email: niel@astro.psu.edu\\
$^2$Max-Planck-Institut f\"ur Extraterrestrische Physik, 85748 Garching, Germany; 
email: grh@xray.mpe.mpg.de}}


\begin{abstract}
Deep surveys of the cosmic \hbox{X-ray} background are reviewed in the context 
of observational progress enabled by the {\it Chandra X-ray Observatory\/} and the
{\it X-ray Multi-Mirror Mission-Newton\/}. The sources found by deep surveys are 
described along with their redshift and luminosity distributions, and the 
effectiveness of such surveys at selecting active galactic nuclei (AGN) 
is assessed. Some key results from deep surveys are highlighted including 
(1) measurements of AGN evolution and the growth of supermassive black holes,
(2) constraints on the demography and physics of high-redshift AGN, 
(3) the \hbox{X-ray} AGN content of infrared and submillimeter galaxies, and 
(4) \hbox{X-ray} emission from distant starburst and normal galaxies.
We also describe some outstanding problems and future prospects for 
deep extragalactic \hbox{X-ray} surveys. 
\end{abstract}

\maketitle


\section{Introduction to Deep Extragalactic \hbox{X-ray} Surveys}

Deep extragalactic surveys have been successful tools in unraveling 
the formation and evolution of cosmic building blocks, including galaxies, 
groups and clusters of galaxies, large-scale structures, and supermassive 
black holes (SMBH). 
Deep surveys act as particularly effective ``time machines'' because fainter 
objects of a given type generally lie at greater distances and therefore 
earlier epochs. Furthermore, deep surveys are often able to probe 
intrinsically less luminous and more typical objects than wide-field, 
shallower surveys. Finally, the fluxes of all detected objects, both 
faint and bright, can be summed and compared with the extragalactic 
background light, which provides an integral census of the emission in 
the corresponding wavelength range. 
The numerous multiwavelength deep surveys centered around the 
Hubble Deep Fields (e.g., Ferguson, Dickinson \& Williams 2000), 
for example, have demonstrated these characteristics impressively. 

This review concentrates on deep extragalactic \hbox{X-ray} surveys 
in the \hbox{0.1--10~keV} band accessible to imaging telescopes. 
Two powerful, currently active \hbox{X-ray} missions, NASA's
{\it Chandra X-ray Observatory\/} (\chandra; 
Weisskopf et~al. 2000) and ESA's {\it X-ray Multi-Mirror Mission-Newton\/} 
(\xmm; Jansen et~al. 2001), have executed a number of deep 
extragalactic \hbox{X-ray} observing programs, which comprise by far the most 
sensitive \hbox{X-ray} surveys ever performed. Building on previous pioneering 
work with the \einstein, \rosat, \asca, and \sax\ missions, these surveys 
resolve the majority of the \hbox{0.1--10~keV} background. 
A substantial amount of multiwavelength follow-up work on the
detected \hbox{X-ray} sources has also been completed. 
It is therefore timely to review the status and 
scientific results of deep extragalactic \hbox{X-ray} surveys. 
In this review, we focus on surveys reaching flux limits of at 
least $5\times 10^{-16}$~erg~cm$^{-2}$~s$^{-1}$ \hbox{(0.5--2~keV)} or 
$1.5\times 10^{-15}$~erg~cm$^{-2}$~s$^{-1}$ \hbox{(2--10~keV)}, corresponding 
to \chandra\ or \xmm\ exposures of $\simgt 75$~ks
(see Figure~1 and Table~1). The equally important
wider-field, shallower \hbox{X-ray} 
surveys are not covered extensively here, 
although they are mentioned when they especially complement deep 
surveys; for example, when we are discussing the evolution of
active galactic nuclei (AGN) and the growth of the SMBH that 
primarily power them.  
Some other in-depth 
reviews of deep extragalactic \hbox{X-ray} surveys and the cosmic
X-ray background (CXRB) are
Fabian \& Barcons (1992), 
Hasinger \& Zamorani (2000), 
Gilli (2004), and 
Brandt et~al. (2005). 
Rosati, Borgani \& Norman (2002a) have recently reviewed \hbox{X-ray} galaxy 
cluster surveys including some of the key results from deep \chandra\ 
and \xmm\ surveys; we shall not repeat this material in detail here. 

In the remainder of this section, we briefly review the history of \hbox{X-ray} 
deep-field research and describe the deepest \chandra\ and \xmm\  
surveys. We also discuss the observed \hbox{X-ray} number counts and
the fraction of the \hbox{0.1--10~keV} background resolved. 
In Section~2 we focus on the basic source types found, the 
observed AGN redshift and luminosity distributions, and the
completeness of AGN \hbox{X-ray} selection. 
Section~3 reviews some key recent results from extragalactic 
X-ray surveys in targeted areas of scientific interest, and 
Section~4 briefly examines some outstanding problems and future 
prospects.

Throughout this review we shall adopt the \wmap\ concordance 
cosmology with 
\hbox{$H_0=70$~km~s$^{-1}$~Mpc$^{-1}$}, 
\hbox{$\Omega_{\rm M}=0.3$}, and 
\hbox{$\Omega_{\Lambda}=0.7$} 
(Spergel et~al. 2003), unless otherwise noted. 


\subsection{Brief Historical Summary of \hbox{X-ray} Deep-Field Research}

The CXRB was discovered by Giacconi et~al. (1962) 
in a rocket flight originally designed to detect
\hbox{X-ray} emission from the Moon; the CXRB was the 
first cosmic background discovered. The 
data showed a strong Galactic \hbox{X-ray} 
source \hbox{(Sco~X-1)} and diffuse emission 
of approximately constant intensity from all 
directions observed during the flight. The 
first all-sky \hbox{X-ray} surveys with \uhuru\ and \arielv\ in the 
1970's revealed a high degree of CXRB 
isotropy leading researchers to 
conclude that the CXRB must be mainly extragalactic. If it were 
comprised of discrete sources, such as AGN (e.g., Setti \& Woltjer 1973), 
the number of sources contributing to the CXRB 
had to be very large: $N>10^6$~sr$^{-1}$ 
(Schwartz 1980). On the other hand, high-quality data 
from \heao1\ showed that the shape of the 3--50~keV CXRB could 
be well fit by an isothermal bremsstrahlung model corresponding 
to an optically thin, hot plasma with $kT\approx 40$~keV 
(Marshall et~al. 1980). This was taken by some to suggest
an origin of the CXRB in a hot intergalactic medium, until
this possibility was ruled out by Compton-distortion 
constraints on the spectrum of the cosmic microwave 
background (see Section~5.1 of Fabian \& Barcons 1992 and 
references therein). 

When sensitive, high angular resolution imaging \hbox{X-ray} observations with 
Wolter telescopes became possible, the discrete nature of the CXRB became 
increasingly clear. Pointed observations of previously known AGN
soon showed that, as a class, they are luminous \hbox{X-ray} emitters 
(e.g., Tananbaum et~al. 1979). Deep \einstein\ observations resolved 
$\approx 25$\% of the \hbox{1--3~keV} CXRB into discrete sources at fluxes 
down to $\approx 3\times 10^{-14}$~erg~cm$^{-2}$~s$^{-1}$, a large fraction 
of which were identified as AGN (Giacconi et~al. 1979). Deep 
\hbox{0.5--2~keV} surveys with \rosat\ to 
limiting fluxes of $\approx 10^{-15}$~erg~cm$^{-2}$~s$^{-1}$ 
were for the first time able to resolve 
the majority ($\approx 75$\%) of the soft CXRB
into discrete sources (e.g., Hasinger et~al. 1993, 1998). 
Extensive optical follow-up spectroscopy identified 
the bulk of these sources as AGN
(e.g., McHardy et~al. 1998; Schmidt et~al. 1998; 
Zamorani et~al. 1999; Lehmann et~al. 2001), 
demonstrating that at least the \hbox{0.5--2~keV} CXRB 
is predominantly due to accretion onto SMBH, 
integrated over cosmic time. The deep \rosat\ surveys 
detected an AGN sky density \hbox{($\approx 780$--870~deg$^{-2}$)} 
larger than at any other wavelength and found evidence for 
luminosity-dependent density evolution of AGN
(Miyaji, Hasinger \& Schmidt 2000), contrary to the pure 
luminosity evolution that had been proposed for both 
optically and \hbox{X-ray} selected AGN (e.g., Boyle et~al. 1993). 

Deep surveys above \hbox{2--3~keV} had to wait considerably 
longer than those at lower energies owing to technological 
challenges. \asca\ performed a number 
of medium-to-deep sky surveys to limiting \hbox{2--10~keV} 
fluxes of \hbox{$10^{-13}$--$5\times 10^{-14}$~erg~cm$^{-2}$~s$^{-1}$}  
(e.g., Georgantopoulos et~al. 1997; Ueda et~al. 1998; 
Cagnoni, della~Ceca \& Maccacaro 1998; 
Ishisaki et~al. 2001). These surveys reached AGN sky 
densities of 10--100~sources~deg$^{-2}$. At higher 
densities these surveys were heavily confused due to the 
limited angular resolution of \asca. An analysis of the 
spatial fluctuations in deep \asca\ images 
probed the \hbox{2--10~keV} source counts down to a flux 
limit of $2\times 10^{-14}$~erg~cm$^{-2}$~s$^{-1}$, 
resolving $\approx 35$\% of the \hbox{2--10~keV} CXRB
(Gendreau, Barcons \& Fabian 1998). Surveys 
in the \hbox{5--10~keV} band were pioneered using \sax, which was
well suited for such work because of its relatively large 
throughput at high energies and its significantly sharper 
point spread function at high energy compared with \asca. 
These observations resolved 20--30\% of the \hbox{5--10~keV} 
CXRB (e.g., Comastri et~al. 2001).

The above \hbox{X-ray} observations were interpreted in the 
context of CXRB population synthesis models based on unified 
AGN schemes (e.g., Madau, Ghisellini \& Fabian 1994; 
Comastri et~al. 1995; Gilli, Salvati \& Hasinger 2001). 
These explain the CXRB spectrum using a 
mixture of obscured and unobscured AGN, folded with the 
corresponding luminosity function and its cosmological evolution. 
According to these models, most AGN spectra are heavily absorbed, 
and about 85\% of the radiation produced by SMBH accretion is obscured
by dust and gas (e.g., Fabian \& Iwasawa 1999). These models predicted a 
significant number of heavily obscured AGN in deep hard \hbox{X-ray} 
fields, which had already been partly found in the \asca\ and \sax\ 
surveys. They also predicted a substantial contribution from 
high-luminosity, obscured \hbox{X-ray} sources (e.g., type~2 quasars), 
which at that time had only scarcely been detected. However, these models 
were far from unique and contained a number of implicit assumptions. 
For instance, evolution of obscuration over cosmic time and the 
dependence of obscuration on intrinsic source luminosity were largely 
free parameters. Nevertheless, these models provided a working framework 
and predictions for deeper surveys, which could be tested with 
\chandra\ and \xmm.



\subsection{Current Deep Surveys with Chandra and XMM-Newton}

The superb Wolter telescopes and charge-coupled device (CCD) detectors 
on \chandra\ and \xmm\ provide deep-survey researchers with the
following: 

\begin{enumerate}

\item
Sensitive imaging spectroscopy from \hbox{$\approx 0.5$--10~keV},
with up to \hbox{50--250} times (depending upon the energy band considered)
the sensitivity of previous \hbox{X-ray} missions. 

\item
X-ray source positions with accuracies of
\hbox{$\approx$~0.3--1$^{\prime\prime}$} (\chandra) and
\hbox{$\approx$~1--3$^{\prime\prime}$} (\xmm). These high-quality 
positions are essential for matching to faint multiwavelength 
counterparts in deep surveys, thereby allowing efficient 
follow-up studies. 

\item
Large source samples (100--600 sources or more, per survey) allowing 
reliable statistical inferences to be drawn about faint extragalactic 
\hbox{X-ray} source populations. 

\end{enumerate}

\noindent
The deep-survey capabilities of \chandra\ and \xmm\ are complementary 
in several important respects. Due to its sub-arcsecond imaging
which provides a small source detection cell, \chandra\ can achieve the 
highest possible \hbox{$\approx$~0.5--8~keV} sensitivity with long 
exposures. The \chandra\ \hbox{0.5--8~keV} background count rate is 
only $\approx 0.2$~count~Ms$^{-1}$~pixel$^{-1}$, and the 
faintest \chandra\ sources detected have 
count rates of $\approx 1$ count every \hbox{2--4} days. 
Even the deepest \chandra\ observations performed to date 
do not suffer from significant source confusion
(Alexander et~al. 2003b), in contrast to
the case for \xmm, where confusion becomes significant for
\hbox{$\simgt 100$--200~ks} exposures.  
\xmm\ has a substantially larger photon collecting area than \chandra, 
allowing efficient \hbox{X-ray} spectroscopy at fluxes above its confusion 
limit. The field of view for \xmm\ is also $\approx 2.5$ times that 
of \chandra.

Table~1 lists the current deep \chandra\ and \xmm\ surveys; these 
21 surveys have a total exposure exceeding 80~days. The most sensitive 
surveys performed by \chandra\ and \xmm, 
the 2.0~Ms \chandra\ Deep Field-North (\hbox{CDF-N}) and 
the 770~ks \xmm\ Lockman Hole field, are shown in Figure~2. 
The surveys in Table~1 have generally been performed in regions of 
sky where 
(1) the extensive requisite multiwavelength supporting data already exist 
and/or some interesting astronomical object is present (e.g., Abell~370, 
3C~295, or the SSA22 ``protocluster''), and
(2) there is little Galactic foreground \hbox{X-ray} absorption
(e.g., Lockman 2004). 
At the flux levels probed by these surveys, even moderate-luminosity 
AGN, similar to Seyfert~1 galaxies in the local universe, can be detected 
to $z\simgt 4$.
%
%
The surveys in Table~1 span a significant range of solid-angle 
coverage; however, they are all ``pencil-beam'' surveys in that
even the widest cover only $\approx 5\times 10^{-5}$ of the sky
(about nine times the solid angle of the full Moon). 


\subsection{Deep-Survey Number Counts and the Fraction of the
Cosmic X-ray Background Resolved}


Based on deep surveys with \chandra\ and \xmm, number-count
relations have now been determined down to 0.5--2, 2--8, and 5--10~keV 
fluxes of about
$2.3\times 10^{-17}$, 
$2.0\times 10^{-16}$, and 
$1.2\times 10^{-15}$~erg~cm$^{-2}$~s$^{-1}$, respectively 
(e.g., Brandt et~al. 2001b; Hasinger et~al. 2001; Cowie et~al. 2002; 
Rosati et~al. 2002b; Moretti et~al. 2003; Bauer et~al. 2004). 
Figure~3 shows the integral number counts in the 0.5--2 and 2--8~keV 
bands. At bright fluxes the integral counts have power-law slopes 
in the range $\alpha_{\rm b}\approx 1.6\pm0.2$, depending on the sample 
selection (compare, e.g., the discussions about bright-end slopes in 
Hasinger et~al. 1998 and Moretti et~al. 2003). Toward fainter
0.5--2 and 2--8~keV fluxes, the integral counts show significant 
cosmological flattening with faint-end slopes of 
$\alpha_{\rm f}\approx 0.4$--0.6 and break fluxes of 
\hbox{$\approx (1$--$2)\times10^{-14}$} and 
\hbox{$\approx (3$--$8)\times10^{-15}$~erg~cm$^{-2}$~s$^{-1}$}, 
respectively. In the 5--10~keV band a flattening has not yet been 
detected; the faint-end number counts continue rising steeply 
($\alpha_{\rm f}\approx 1.2$--1.4) indicating that a significant fraction 
of the 5--10~keV CXRB remains unresolved (e.g., Rosati et~al. 2002b). 

There is some evidence for field-to-field variations of the number counts. 
Such variations are expected at some level due to ``cosmic variance'' 
associated with large-scale structures that have been detected
in the \hbox{X-ray} sky (e.g., Barger et~al. 2003a; 
Gilli et~al. 2003, 2005; Yang et~al. 2003). 
For example, though the \hbox{CDF-N} and \chandra\ Deep Field-South (CDF-S) 
number counts agree in the 0.5--2~keV band 
and at bright 2--8~keV fluxes, there is up to 
$\approx 3.9\sigma$ disagreement for 2--8~keV fluxes below
$\approx 1\times 10^{-15}$~erg~cm$^{-2}$~s$^{-1}$ 
(Cowie et~al. 2002; Rosati et~al. 2002b; Bauer et al. 2004). 
The number counts for the shallower Lockman Hole (Hasinger et~al. 2001) 
and Lynx (Stern et~al. 2002a) fields agree with those for the \chandra\ Deep 
Fields to within statistical errors, whereas those for the SSA13 
(Mushotzky et~al. 2000) field appear to be $\approx 40$\% higher in the 
2--8~keV band (see Tozzi et~al. 2001). An extensive comparison of 
field-to-field number counts by Kim et~al. (2004) finds little evidence 
for cosmic variance at \hbox{0.5--2~keV} (\hbox{2--8~keV}) flux 
levels of $\approx 10^{-15}$--$10^{-13}$~erg~cm$^{-2}$~s$^{-1}$
($\approx 10^{-14}$--$10^{-12}$~erg~cm$^{-2}$~s$^{-1}$)
in $\approx 5$--125~ks \chandra\ observations. 

The deepest \rosat\ surveys resolved $\approx 75$\% of the 0.5--2~keV 
CXRB into discrete sources, the major uncertainty in the resolved fraction
being the absolute flux level of the CXRB (at low energies it is 
challenging to separate the CXRB from Galactic emission; see
McCammon \& Sanders 1990). Deep \chandra\ and \xmm\ surveys 
have now increased this resolved fraction to $\approx 90$\% 
(e.g., Moretti et~al. 2003; Bauer et~al. 2004; Worsley et~al. 2004). 
Above 2~keV the situation is complicated by the fact that 
the \heao1\ background spectrum (Marshall et~al. 1980), used as a 
reference for many years, has a $\approx 30\%$ lower normalization than several 
earlier and later background measurements (see, e.g., Moretti et~al. 2003). 
Recent determinations of the background spectrum with \rxte\ (Revnivtsev et~al. 2003) 
and \xmm\ (De~Luca \& Molendi 2004) strengthen the consensus for a 30\% higher 
normalization, indicating that many past resolved fractions 
above 2~keV must be scaled down correspondingly. 
Additionally, X-ray telescopes typically have a 
large sensitivity gradient across the broad 2--10~keV band.
A recent investigation by Worsley et~al. (2004), dividing the \hbox{CDF-N}, 
\hbox{CDF-S}, and \xmm\ Lockman Hole field into finer energy bins, concludes 
that the resolved fraction drops from $\approx 80$--90\% at 2--6~keV to 
$\approx 50$--70\% at 6--10~keV. This is consistent with expectations from 
the 5--10~keV number counts (see above). 
In the critical 10--100~keV band, where most of the CXRB energy 
density resides, only a few percent of the background has been 
resolved (e.g., Krivonos et~al. 2005). 

Multiwavelength identification studies indicate that
most \hbox{($\simgt 70$\%)} of the \hbox{X-ray} sources found in deep 
\chandra\ and \xmm\ surveys are AGN (see Section~2.1 for further discussion). 
The observed AGN sky density in the deepest \hbox{X-ray} surveys, 
the \chandra\ Deep Fields, is a remarkable $\approx 7200$~deg$^{-2}$
(e.g., Bauer et~al. 2004). This exceptional effectiveness at finding 
AGN arises largely because \hbox{X-ray} selection (1) has reduced 
absorption bias, (2) has minimal dilution by host-galaxy starlight, and 
(3) allows concentration of intensive optical spectroscopic follow-up 
upon high-probability AGN with faint optical counterparts (i.e., it 
is possible to probe further down the luminosity function);
see Section~2.4 and Mushotzky (2004) for further details on the 
effectiveness of AGN \hbox{X-ray} selection. The AGN sky density 
from the \chandra\ Deep Fields exceeds that found at
any other wavelength and is 10--20 times higher than that found in the 
deepest optical spectroscopic surveys (e.g., Wolf et~al. 2003; 
Hunt et~al. 2004); only ultradeep optical variability studies 
(e.g., Sarajedini, Gilliland \& Kasm 2003) may be generating comparable 
AGN sky densities. 


\section{Properties of the Sources Found and Missed by 
Deep Extragalactic X-ray Surveys}

A broad diversity of source types is found in deep \chandra\ and 
\xmm\ surveys. This is apparent in even basic flux-flux plots such as that
shown in Figure~4; at the faintest \hbox{X-ray} flux levels in the 
\chandra\ Deep Fields, the extragalactic sources range in optical flux 
by a factor of $\simgt 10,000$. 


\subsection{Source Classification Challenges}

Classification of deep X-ray survey
sources is challenging for several reasons. 
First, many of the X-ray detected AGN are 
simply too faint for straightforward
optical spectroscopic identification 
even with \hbox{8--10~m} class telescopes 
(note the multitude of small dots in 
Figure~4). Intensive optical identification 
programs on deep \chandra\ and \xmm\ fields typically have 
\hbox{50--70\%} redshift completeness at best
(e.g., Barger et~al. 2003a; Szokoly et~al. 2004), and some of the 
obtained spectra that do yield redshifts are of insufficient
signal-to-noise for reliable optical classification work. 
Furthermore, many of the \hbox{X-ray} sources have modest optical 
luminosities, often due to obscuration. Thus starlight 
from their host galaxies can make a substantial 
diluting contribution to the flux measured in a ground-based 
spectroscopic aperture, plausibly ``overwhelming''
subtle nuclear spectral features from an obscured or
low-luminosity AGN (e.g., Moran, Filippenko, \& Chornock 2002). 
Finally, another challenge is an apparent ``schism'' between optical 
(type~1 vs. type~2) and \hbox{X-ray} (unobscured vs. obscured) schemes of
classification; not all \hbox{X-ray} obscured AGN have type~2 optical spectra, and 
not all AGN with type~1 optical spectra are X-ray unobscured (e.g., Matt 2002). 

\subsection{Basic Source Types}

Most of the \hbox{X-ray} sources found in deep surveys are 
AGN. Considering \hbox{X-ray}, optical, and multiwavelength 
information, the types of AGN found include the following: 

\begin{enumerate}

\item
{\it Unobscured AGN.\/} 
Blue, broad-line AGN are found that do not show
signs of obscuration at either \hbox{X-ray} or optical/UV wavelengths. They
cover a broad range of redshift \hbox{($z\approx 0$--5)}, and they
comprise a significant fraction of the brightest \hbox{X-ray} sources
(e.g., Barger et~al. 2003a; Szokoly et~al. 2004). At 
$z\simgt 1.5$ they also comprise a substantial fraction of all \hbox{X-ray} 
sources with spectroscopic identifications, certainly in part because these
objects are the most straightforward to identify spectroscopically. 

\item
{\it Obscured AGN with clear optical/UV AGN signatures.\/} 
Some objects showing \hbox{X-ray} evidence for obscuration have clear AGN 
signatures in their rest-frame optical/UV spectra. Notably, such AGN 
can have both type~1 and type~2 optical/UV classifications
(e.g., Matt 2002). Most of these objects have $z\simlt 1.5$. 

\item
{\it Optically faint X-ray sources.\/} 
These sources have $I>24$--25 and usually cannot be 
identified spectroscopically (see Figure~4). Many, 
however, appear to be luminous, obscured 
AGN at \hbox{$z\approx 1$--4} when their \hbox{X-ray} properties, 
optical photometric properties (including photometric redshifts), and 
multiwavelength properties are considered
(e.g., Alexander et~al. 2001; Barger et~al. 2003a; Fiore et~al. 2003; 
Treister et~al. 2004; Zheng et~al. 2004; Mainieri et~al. 2005). 
Thus, many of these objects likely represent an extension of the 
previous class to higher redshifts and fainter optical magnitudes. 
Some have no optical detections at all, even in 
deep images, and are termed extreme \hbox{X-ray}/optical ratio
sources (EXOs); most of these are detected in near-infrared images
(Koekemoer et~al. 2004; Mignoli et~al. 2004).
EXOs usually can be plausibly explained as 
$z\approx 1.5$--5 obscured AGN in dusty or evolved hosts, although
a minority may lie at $z\simgt 7$ where the intergalactic medium
absorbs essentially all of the observed-frame optical emission. 

\item
{\it X-ray bright, optically normal galaxies (XBONGs).\/} 
XBONGs are early-type galaxies at \hbox{$z\approx 0.05$--1}
that have \hbox{X-ray} luminosities 
($\approx 10^{41}$--$10^{43}$~erg~s$^{-1}$), 
\hbox{X-ray} spectral shapes, and X-ray-to-optical flux ratios 
suggesting AGN activity of moderate strength 
(e.g., Comastri et~al. 2005). However, AGN emission lines and 
non-stellar continua are not apparent in optical spectra.
Some XBONGs have \hbox{X-ray} spectra suggesting the 
presence of obscuration whereas others do not (e.g., Severgnini et~al. 2003). 
The nature of XBONGs remains somewhat mysterious. Many and
perhaps most may just be normal
Seyfert galaxies where dilution by host-galaxy light 
hinders optical detection of the AGN 
(e.g., Moran et~al. 2002; Severgnini et~al. 2003), but 
some have high-quality follow-up observations and appear 
to be truly remarkable objects (e.g., Comastri et~al. 2002). These 
``true'' XBONGs may be (1) AGN with inner radiatively inefficient accretion 
flows (Yuan \& Narayan 2004), (2) AGN that suffer from heavy obscuration 
covering a large solid angle ($\approx 4\pi$~sr), so that optical emission-line 
and ionizing photons cannot escape the nuclear region (e.g., Matt 2002),
or, in some cases, (3) BL Lac-like objects (e.g., Brusa et~al. 2003). 
XBONGs appear related to ``optically dull galaxies'' (e.g., Elvis et~al. 1981) 
and ``elusive AGN'' (e.g., Maiolino et~al. 2003) studied in the local universe. 

\end{enumerate}

\noindent
Morphological studies show that the AGN from deep \hbox{X-ray} surveys are 
generally hosted by galaxies with significant bulge components,
and they do not show evidence for enhanced merging or interaction
activity relative to normal field galaxies (e.g., Grogin et~al. 2005). 
Many of the non broad-line AGN have the rest-frame colors of 
evolved, bulge-dominated galaxies, and there is little evolution of
these colors from $z\approx 0$--2 (e.g., Barger et~al. 2003a; 
Szokoly et~al. 2004).
Most AGN from deep surveys are ``radio quiet'' in the sense that the 
ratio, $R$, of their rest-frame 5~GHz and 4400~\AA\ flux densities is $R<10$
(e.g., Bauer et~al. 2002b). 

In addition to AGN, several other types of objects are found in 
deep extragalactic \hbox{X-ray} surveys. These include the following: 

\begin{enumerate}

\item
{\it Starburst and normal galaxies.\/}
At the faintest measured \hbox{X-ray} flux levels 
(0.5--2~keV fluxes below $\approx 5\times 10^{-16}$~erg~cm$^{-2}$~s$^{-1}$;
see Figure~3), a significant and rising (up to 30--40\%) fraction of the detected 
sources appears to be starburst and normal galaxies at \hbox{$z\approx 0.1$--1.5}
(e.g., Hornschemeier et~al. 2003; Bauer et~al. 2004). 
These galaxies are discussed in more detail in Section~3.4. 

\item
{\it Groups and clusters of galaxies.\/}
Groups and low-luminosity clusters of galaxies at $z\approx 0.1$--1 are 
detected as extended, soft \hbox{X-ray} sources in deep surveys
(e.g., Bauer et~al. 2002a; Giacconi et~al. 2002; Rosati et~al. 2002a). 
Their \hbox{X-ray} luminosities ($L_{\rm X}\approx 10^{41.5}$--$10^{43}$~erg~s$^{-1}$), 
basic \hbox{X-ray} spectral properties ($kT\simlt 3$~keV), and sizes
appear broadly consistent with those of nearby groups and low-luminosity 
clusters. The surface density of extended 
X-ray sources is $\approx 100$--260~deg$^{-2}$ at a limiting 
0.5--2~keV flux of $\approx 3\times 10^{-16}$~erg~cm$^{-2}$~s$^{-1}$; 
no strong evolution in the \hbox{X-ray} luminosity function of clusters 
is needed to explain this value.

\item
{\it Galactic stars.\/}
Some stars are detected in the high Galactic latitude fields targeted
by deep surveys. These are typically of type G, K, and M, and their
X-ray emission is attributed to magnetic reconnection flares. 
The observed \hbox{X-ray} emission constrains the decay of late-type
stellar magnetic activity on timescales of $\approx 3$--11~Gyr
(e.g., Feigelson et~al. 2004). One of the stars detected in the
Lynx field has notably hard \hbox{X-ray} emission, and Stern et~al. (2002a)
propose this is a binary system where accretion powers the \hbox{X-ray}
emission. 

\end{enumerate}

Figure~5 shows the source classifications in the {\it Hubble\/} Deep Field-North
\hbox{(HDF-N)} and the {\it Hubble\/} Ultradeep Field \hbox{(UDF)}. These lie 
near the centers of the \hbox{CDF-N} (see Figure~2a) and \hbox{CDF-S}, respectively, 
and thus have the most sensitive \hbox{X-ray} coverage available. Despite the many 
intensive studies of the \hbox{HDF-N} prior to the acquisition of the \chandra\ 
data, several new AGN were identified in the \hbox{HDF-N} based upon their observed 
X-ray emission (e.g., the three XBONGs; see Brandt et~al. 2001a). 

Contributions to the total \hbox{X-ray} number counts from a few of the 
source classes mentioned above are shown in Figure~3. Bauer et~al. (2004)
provide detailed decompositions of the number counts by source class, 
luminosity range, and estimated amount of \hbox{X-ray} absorption. 


\subsection{AGN Redshift and Luminosity Distributions}

Most spectroscopically identified AGN in deep \hbox{X-ray} surveys have 
$z\approx 0$--2, although a significant minority have \hbox{$z\approx 2$--5}. 
This is partly due to spectroscopic incompleteness bias for faint AGN 
at $z\simgt 2$, as is apparent by noting the systematic progression of 
symbol colors and sizes in Figures~4 and 6. The observed 
redshift distribution is concentrated at 
significantly lower redshifts than those predicted by pre-\chandra\ 
population-synthesis models of the CXRB 
(see Section~1.1). Application of photometric and 
other redshift estimation techniques to optically faint \hbox{X-ray} sources 
partially mitigates this discrepancy, as many of these sources are estimated 
to lie at $z\approx 1$--4 (see Section~2.2). However, as described 
further in Section~3.1, there remains a low-redshift enhancement of AGN relative 
to expectations from pre-\chandra\ population-synthesis models. About 60\% 
of the 2--8~keV CXRB arises at $z<1$. 

The observed redshift distributions from deep \hbox{X-ray} surveys also 
show significant ``spikes'' in narrow redshift ranges from $z\approx 0.5$--2.5
(e.g., Barger et~al. 2003a; Gilli et~al. 2003, 2005); spikes at
corresponding redshifts are also seen for sources selected at 
other wavelengths. These are associated with 
X-ray source clustering in large-scale sheet-like structures with sizes
of $\simgt 5$--10~Mpc. These structures are likely a cause of the apparent
cosmic variance mentioned in Section~1.3, and further studies of AGN clustering
in deep \hbox{X-ray} surveys should determine if AGN fueling depends significantly
upon large-scale environment. 

The sources creating most of the 0.1--10~keV CXRB have \hbox{X-ray} 
luminosities of $10^{42}$ to a few times $10^{44}$~erg~s$^{-1}$ 
(see Figure~6), comparable to those of local Seyfert~1 
galaxies (e.g., NGC~3783, NGC~4051, and NGC~5548)  
and lower luminosity quasars (e.g., I~Zwicky~1).
The fraction of AGN showing evidence for significant \hbox{X-ray} obscuration 
drops with increasing luminosity from 
\hbox{$\approx 60$\%} at $10^{42}$~erg~s$^{-1}$
to $\approx 30$\% at $10^{45}$~erg~s$^{-1}$ (e.g., Ueda et~al. 2003;
Szokoly et~al. 2004). A number of X-ray obscured quasars have been found 
in deep surveys (e.g., Norman et~al. 2002; Stern et~al. 2002b; 
Barger et~al. 2003a; Szokoly et~al. 2004). These generally 
have \hbox{X-ray} luminosities of $10^{44}$--$10^{45}$~erg~s$^{-1}$, just above
those of powerful Seyfert galaxies. Some are optically faint
or have limited rest-frame spectral coverage (e.g., low-order Balmer
lines that can penetrate several magnitudes of extinction are not covered) 
so that a type~2 optical classification is difficult to prove
rigorously. Obscured quasars create $\approx 10$\% of the 0.1--10~keV 
CXRB.


\subsection{AGN Selection Completeness}

Are there significant numbers of luminous AGN that are not found 
even in the deepest \hbox{X-ray} surveys? 
This could be the case if there is a population of AGN
that is luminous at non-X-ray wavelengths but is \hbox{X-ray} weak,
perhaps due to an intrinsic inability to produce \hbox{X-rays}. 
For example, such AGN could lack accretion-disk coronae, the
structures putatively responsible for creating much 
of the emission above $\approx 0.5$--1~keV. 
Such intrinsically \hbox{X-ray} weak AGN appear to be rare, 
in line with the dogma that \hbox{X-ray} emission is a universal property 
of AGN, although some may exist (e.g., Brandt, Laor \& Wills 2000; 
Risaliti et~al. 2003; Brandt, Schneider \& Vignali 2004a; 
Leighly, Halpern \& Jenkins 2004). 

Alternatively, strong \hbox{X-ray} absorption could render a luminous AGN 
X-ray weak, even if it were intrinsically producing X-rays at a nominal 
level. Such absorption is seen in ``Compton-thick'' AGN, which 
comprise $\approx 40$\% or more of AGN in the local universe 
(e.g., Risaliti, Maiolino \& Salvati 1999; Comastri 2004). 
Compton-thick AGN are absorbed by column densities of 
$N_{\rm H}\gg 1.5\times 10^{24}$~cm$^{-2}$, so that the optical
depth to electron scattering is $\tau\gg 1$ (for comparison, 
the column density through a person's chest is 
$N_{\rm H}\approx 10^{24}$~cm$^{-2}$). Within such 
thick absorption, direct line-of-sight \hbox{X-rays} are 
effectively destroyed via the combination of Compton scattering and 
photoelectric absorption, even at high energies of 10--200~keV. 
Compton-thick AGN are thus only visible via indirect \hbox{X-rays} 
that are $\approx 50$--150 times weaker (e.g., Comastri 2004); 
these reach an observer via a less-obscured path by 
``reflecting'' off neutral material and ``scattering'' off diffuse
ionized material.\footnote{In some ``translucent'' cases, where the 
column density is only a few times $10^{24}$~cm$^{-2}$, direct 
``transmission'' X-rays from a Compton-thick AGN may become visible 
above rest-frame energies of $\approx 10$~keV.} 

Some well-known, nearby Compton-thick AGN are shown in Figure~6 
along the left-hand side of the plot. They are plotted at the
level of their {\it observable\/} \hbox{X-ray} luminosities; their
absorption-corrected luminosities would be $\approx 50$--150 
times higher. Note that NGC~1068, NGC~4945, and the Circinus Galaxy 
would become undetectable in the 2.0~Ms CDF-N if placed at 
$z\simgt 0.1$--0.5, whereas NGC~6240 and Mrk~231 would become 
undetectable at $z\simgt 2$. It thus appears likely that many 
Compton-thick AGN remain undetected in even the deepest 
X-ray surveys to date. 
Only a small number of Compton-thick AGN have been isolated among 
the currently detected \hbox{X-ray} sources in deep surveys. However, 
since Compton-thick AGN have a diversity of complex \hbox{X-ray} spectral 
shapes and are expected to be \hbox{X-ray} faint, additional ones
could be residing among the currently detected 
sources with limited photon statistics. More refined
searches for Compton-thick AGN among the detected \hbox{X-ray} sources are 
underway, utilizing characteristic \hbox{X-ray} signatures 
(e.g., strong iron~K$\alpha$ lines) and 
new multiwavelength data
[e.g., {\it Spitzer Space Telescope\/} (\spitzer) measurements of 
re-radiated infrared ``waste heat''].

If there is indeed a large population of luminous AGN that has
eluded detection in the deepest \hbox{X-ray} surveys, this population 
also appears to have mostly eluded AGN searches at other 
wavelengths. There are only a few secure AGN in the \chandra\ 
Deep Fields, for example, that have not been detected in 
\hbox{X-rays} (see Section~4 of Bauer et~al. 2004). These include 
an optically selected, narrow-line AGN at $z=2.445$ (Hunt et~al. 2004), 
a radio-bright ($\approx 6$~mJy at 1.4~GHz) wide-angle-tail 
source at \hbox{$z\approx 1$--2} (e.g., Snellen \& Best 2001), and 
perhaps a couple of AGN selected optically in the COMBO-17 
survey (Wolf et~al. 2003).\footnote{Although the wide-angle-tail
radio source is not included in the CDF-N \hbox{X-ray} source catalogs of 
Alexander et~al. (2003b), manual analysis of the \chandra\ data 
at the AGN position indicates a likely detection.}
Sarajedini, Gilliland \& Kasm (2003) also reported some 
galaxies with optically variable nuclei in the \hbox{HDF-N} that 
have not been detected in X-rays. Although some of these may be
nuclear supernovae or statistical outliers, some are plausibly
low-luminosity AGN with \hbox{X-ray} fluxes that lie below the
current \hbox{X-ray} sensitivity limit. 

Another way to assess AGN selection completeness in deep \hbox{X-ray} 
surveys is to consider ``book-keeping'' arguments: can the observed 
sources explain the total CXRB intensity, and can all
the observed accretion account for the local density of SMBH? 
The answer to the first question is
``no'' according to the \hbox{5--10~keV} number counts and 
source-stacking analyses, as described in Section~1.3. 
Worsley et~al. (2004) propose that the sky density of obscured, 
X-ray undetected AGN may be \hbox{$\approx 2000$--3000~deg$^{-2}$} 
or higher. 
The answer to the second question is plausibly ``yes'' but with 
some uncertainty; this is discussed further in Section~3.1. 


\section{Some Key Results from Deep Extragalactic \hbox{X-ray} Surveys}


\subsection{X-ray Measurements of AGN Evolution and the Growth of 
Supermassive Black Holes}

Optical studies of AGN evolution have historically focused on luminous quasars
(e.g., Hewett \& Foltz 1994; Osmer 2004). These have been known to evolve 
positively with redshift since $\approx 1968$ (Schmidt 1968), having a comoving 
space density at $z\approx 2$ that is $\simgt 100$ times higher than at 
$z\approx 0$. Pure luminosity evolution (PLE) models provide acceptable fits 
to large optically selected samples such 
as $\approx 16,800$ luminous AGN from a combination of 
the recent 2dF and 6dF surveys out to \hbox{$z=2.1$} (Croom et~al. 2004). 
At \hbox{$z\simgt 2.7$}, the space density of luminous quasars
selected in wide-field optical multicolor and emission-line surveys shows 
a strong decline with redshift (e.g., Schmidt, Schneider \& Gunn 1995; 
Fan et~al. 2001). Deep optical AGN surveys, such as COMBO-17, 
have recently discovered significant numbers of moderate-luminosity AGN 
(with $M_{\rm B}>-23$) at \hbox{$z\approx 1$--4}
(Wolf et~al. 2003). Like luminous quasars, the moderate-luminosity AGN 
found in these surveys also appear to peak in comoving space density 
around $z\approx 2$. Although a careful merging of the COMBO-17 data 
with a large sample of higher luminosity AGN has yet to be published, there 
are hints that the comoving space density of moderate-luminosity AGN
peaks at smaller redshift (L. Wisotzki, pers. comm.). 

As described in Sections~1.3 and 2.3, 
deep \hbox{X-ray} surveys efficiently select 
AGN significantly less luminous than those found in optical surveys out to 
high redshift. Deep plus wide \hbox{X-ray} surveys can therefore cover an 
extremely broad range of luminosity. Contrary to the PLE model for 
optically selected luminous quasars, the moderate-luminosity 
AGN selected in the \hbox{0.5--2~keV} 
and \hbox{2--10~keV} bands require luminosity-dependent density 
evolution (LDDE; e.g.,
Miyaji, Hasinger \& Schmidt 2000; 
Cowie et~al. 2003; 
Fiore et~al. 2003; 
Steffen et~al. 2003;
Ueda et~al. 2003; 
Barger et~al. 2005; 
Hasinger 2005; 
Hasinger, Miyaji \& Schmidt 2005). 
Figure~7 shows the \hbox{X-ray} luminosity function based on $\approx 950$ AGN 
selected in the \hbox{0.5--2~keV} band from deep \chandra\ and \xmm\ surveys
as well as deep plus wide \rosat\ surveys (Hasinger, Miyaji \& Schmidt 2005). 
Strong positive evolution from \hbox{$z\approx 0$--2} is 
only seen at high luminosities; lower luminosity AGN 
evolve more mildly. The evolutionary behavior measured 
in the 2--10~keV band is similar. These results are robust to incompleteness 
of the spectroscopic follow up, although clearly they are still dependent 
upon the completeness of AGN \hbox{X-ray} selection (see Section~2.4). 
At a basic level, LDDE is not unexpected as simple PLE and pure density
evolution (PDE) models have physical difficulties (see, e.g., 
Weedman 1986 and Peterson 1997). Simple PLE models tend to overpredict 
the number of $\simgt 10^{10}$~M$_\odot$ black holes in the local universe, 
whereas simple PDE models tend to overpredict the local space density 
of quasars and the CXRB intensity. 

Figure~8 shows estimates of the comoving space density of AGN in different 
\hbox{X-ray} luminosity ranges as a function of redshift. 
Figure~8a has been constructed for the \hbox{0.5--2~keV} band
using the Hasinger, Miyaji \& Schmidt (2005) sample. 
Figure~8b is for the \hbox{2--10~keV} band, utilizing a combination 
of \chandra, \asca, and \heao1\ surveys with 247 AGN in 
total (Ueda et~al. 2003).
These plots illustrate that (1) the AGN peak space density moves to smaller 
redshift with decreasing luminosity, and (2) the rate of evolution from the 
local universe to the peak redshift is slower for less-luminous AGN. 
It appears that SMBH generally grow in an ``anti-hierarchical'' fashion:
while the $10^{7.5}$--$10^{9}~M_\odot$ SMBH in rare, luminous AGN could grow 
efficiently at $z\approx 1$--3, the $10^{6}$--$10^{7.5}~M_\odot$ SMBH in 
more-common, less-luminous AGN had to wait longer to grow ($z\simlt 1.5$). 
In the \hbox{0.5--2~keV} band the sensitivity and statistics are good 
enough to detect a clear decline of the luminous AGN space density toward 
higher redshifts (also see Silverman et~al. 2005); such 
a high-redshift decline is also hinted at in some
\hbox{2--10~keV} band analyses (Fiore et~al. 2003; Steffen et~al. 2003). 

The AGN luminosity function can be used to predict the masses of remnant 
SMBH in galactic centers. This is done using the ingenious So\l tan (1982) 
continuity argument, adopting an AGN
mass-to-energy conversion efficiency and  
bolometric correction factor.
The local mass density of SMBH in dormant quasar 
remnants originally predicted by So\l tan (1982) was 
\hbox{$\rho_\bullet > 0.47 \epsilon_{0.1}^{-1} \times 10^5~M_\odot$~Mpc$^{-3}$}, 
where $\epsilon_{0.1}$ is the mass-to-energy conversion efficiency of 
the accretion process divided by 0.1. For a Schwarzschild black hole, 
$\epsilon$ is expected to be 0.054 or larger, depending upon the 
accretion-disk torque at the marginally stable orbit 
around the black hole (e.g., Agol \& Krolik 2000). 
For a rapidly rotating Kerr black hole, $\epsilon$ can be as high as 
$\approx 0.36$. More recent determinations of $\rho_\bullet$
from optical quasar luminosity functions are around 
\hbox{$2 \epsilon_{0.1}^{-1} \times 10^5~M_\odot$ Mpc$^{-3}$} 
(e.g., Chokshi \& Turner 1992; Yu \& Tremaine 2002). Estimates from 
the CXRB spectrum, including obscured accretion power, originally obtained 
even larger values: \hbox{6--9} (Fabian \& Iwasawa 1999) or \hbox{8--17} 
(Elvis, Risaliti \& Zamorani 2002) in the above units.
However, these were derived assuming that the evolution of 
moderate-luminosity AGN is the same as that of quasars 
(i.e., that moderate-luminosity peak in number density at $z\approx 2$),
and they need to be revised downward by a factor of $\approx 3$ in 
light of the currently observed evolution of CXRB sources (Fabian 2004).
Values derived from the infrared band (Haehnelt \& Kauffmann 2001) or 
multiwavelength observations (Barger et~al. 2001b) are high (\hbox{8--9}
in the above units). Arguably the 
most reliable determination comes from an 
integration of the Ueda et~al. (2003) hard \hbox{X-ray} luminosity function 
that includes a revised bolometric correction (which ignores infrared emission
to avoid the double counting of luminosity) and a plausible
correction for missed Compton-thick AGN: 
\hbox{$\rho_\bullet\approx 3.5 \epsilon_{0.1}^{-1} \times 10^5~M_\odot$~Mpc$^{-3}$}
(Marconi et~al. 2004). 

The SMBH masses measured in local galaxies are correlated with the 
velocity dispersions and luminosities of their host bulges 
(e.g., Kormendy \& Gebhardt 2001; Ferrarese \& Ford 2005; and 
references therein). Using these correlations with 
velocity dispersion functions or luminosity functions for
local galaxies, the total SMBH density in galactic bulges can be estimated. 
Scaled to the same assumption for the Hubble constant (see Section~1), 
several recent papers arrive at different values of $\rho_\bullet$, mainly 
depending on assumptions about the SMBH-galaxy correlations. 
For example, Yu \& Tremaine (2002) derive 
\hbox{$(2.9\pm 0.5)\times 10^5~M_\odot$~Mpc$^{-3}$},
whereas Marconi et~al. (2004) derive
$(4.6^{+1.9}_{-1.4})\times 10^5~M_\odot$~Mpc$^{-3}$. 
These values are plausibly consistent with the current best estimates
of accreted mass density for $\epsilon\approx 0.1$. 
Marconi et~al. (2004) and Shankar et~al. (2004) also 
demonstrate that the observed accretion can plausibly explain the 
measured distribution function of local SMBH masses. 


\subsection{X-ray Constraints on the Demography and Physics 
of High-Redshift \boldmath$(z>4)$ AGN}

Deep \hbox{X-ray} surveys can find $z>4$ AGN that are $\approx 10$--30 
times less luminous than the quasars found in wide-field optical surveys
(see Figure~6). Such moderate-luminosity AGN are much more numerous and 
thus more representative of the AGN population than the rare, highly 
luminous quasars. Furthermore, unlike the rest-frame ultraviolet light 
sampled in ground-based surveys of $z>4$ AGN, \hbox{X-ray} surveys 
suffer from progressively less absorption bias as higher redshifts are 
surveyed. At $z>4$, hard $\approx 2$--50~keV rest-frame \hbox{X-rays} 
are observed. 

Optical spectroscopic follow-up of moderate-luminosity \hbox{X-ray} detected AGN 
at $z>4$ is challenging, as such objects are expected to have $I$-band 
magnitudes of \hbox{$\approx 23$--27} (provided they are not at $z\simgt 6.5$, 
where they largely disappear from the $I$ bandpass because of intergalactic 
absorption). Nevertheless, significant constraints on the sky density of such 
objects have been set via large-telescope spectroscopy and Lyman-break selection
(e.g., Alexander et~al. 2001; Barger et~al. 2003b; Cristiani et~al. 2004; 
Koekemoer et~al. 2004; Wang et~al. 2004c). 
The ``bottom line'' from these demographic studies 
is that the sky density of $z>4$ AGN is $\approx 30$--150~deg$^{-2}$ at a 
0.5--2~keV flux limit of $\approx 10^{-16}$~erg~cm$^{-2}$~s$^{-1}$;
for comparison, the Sloan Digital Sky Survey finds a sky density of 
$z=4$--5.4 quasars of $\approx 0.12$~deg$^{-2}$ at an $i$-magnitude
limit of $\approx 20.2$ (e.g., Schneider et~al. 2003). 
Combined demographic constraints from 
\hbox{X-ray} and wide-field optical surveys indicate that the AGN 
contribution to reionization at $z\approx 6$ is small. Significantly 
better source statistics are needed to refine constraints on the 
faint end of the AGN \hbox{X-ray} luminosity function at \hbox{$z>4$},
as only six moderate-luminosity, X-ray selected AGN at \hbox{$z>4$} are 
presently published. The requisite improvements are expected as 
follow-up studies of the surveys in Table~1 
progress. The current \hbox{X-ray} constraints at $z>4$ are plausibly consistent
with the optical luminosity function from COMBO-17 (Wolf et~al. 2003). 

Once $z>4$ AGN have been identified, broad-band spectral energy 
distribution analyses and \hbox{X-ray} spectral fitting provide 
information on their accretion processes and environments. The currently 
available data suggest that $z>4$ AGN spanning a broad luminosity 
range are accreting in basically the same mode as AGN in the local 
universe (e.g., Vignali et~al. 2002, 2003; 
Brandt et~al. 2004b). The \hbox{X-ray} power-law photon 
indices of $z>4$ and low-redshift AGN appear consistent, and their 
X-ray-to-optical flux ratios also agree after allowing for 
luminosity effects. The basic consistency observed provides confidence 
that \hbox{X-ray} surveys should remain an effective way to find AGN at
the highest redshifts. 

To constrain even lower luminosity AGN populations at $z>4$, source-stacking 
analyses have been employed. These search for an average \hbox{X-ray} signal from 
a set of high-redshift sources whose individual members lie below the 
single-source \hbox{X-ray} detection limit. The most sensitive 
$z>4$ source-stacking analyses
to date have employed samples of $\approx 250$--1700 Lyman break galaxies 
(e.g., Lehmer et~al. 2005; $B$, $V$, and $i$ dropouts) and
$\approx 100$ Ly$\alpha$ emitters (e.g., Wang et~al. 2004b). 
The resulting average \hbox{X-ray} detections and upper limits are consistent 
with any \hbox{X-ray} emission from these objects arising from 
stellar processes; thus no AGN emission has clearly been detected. These 
average constraints at luminosities below those that can be probed by 
single-source analyses further limit the contribution that AGN could 
have made to reionization at $z\approx 6$ and point to stars as the
culprit. A complementary average constraint, derived by considering 
the unresolved component of the \hbox{0.5--2~keV} CXRB, provides 
additional evidence that AGN and lower mass black holes did not dominate 
reionization (Dijkstra, Haiman \& Loeb 2004). 


\subsection{X-ray Constraints on AGN in Infrared and Submillimeter Galaxies}

Deep \hbox{X-ray} surveys arguably allow the best assessment of the AGN content 
of distant infrared and submillimeter galaxies
in which much of cosmic star formation took place. AGN are 
expected to be significant emitters at these observed wavelengths, owing 
to the reprocessing of their strong shorter wavelength 
emission by omnipresent dust and gas in their nuclei
(e.g., Almaini, Lawrence \& Boyle 1999). Overall, however, the 
current evidence points to a picture where AGN emission is 
dominated by that from star-formation processes. 

In the infrared band, only a minority of 15~$\mu$m galaxies are
identified with \hbox{X-ray} detected AGN; the majority of 15~$\mu$m/\hbox{X-ray}
matches appear to be starburst galaxies (e.g., Alexander et~al. 2002; 
Fadda et~al. 2002). AGN nuclear emission 
contributes \hbox{$\approx 3$--5\%} or less 
of the total background in most infrared bands, 
the rest coming from starburst and normal galaxies
(e.g., Silva, Maiolino \& Granato 2004). 
The fractional AGN contribution 
could be significantly underestimated if there is a large population of 
infrared-emitting AGN missed by deep \hbox{X-ray} surveys (see Section~2.4). 
For example, the elusive Compton-thick AGN may dominate the AGN 
contribution to the infrared background, although inclusion of these 
probably still leaves AGN as minority contributors to the 
infrared background. \spitzer\ should be 
able to detect the putative Compton-thick AGN missed by
deep \hbox{X-ray} surveys (e.g., Rigby et~al. 2004; Treister et~al. 2004).
However, unambiguously separating these from the numerous starburst 
galaxies detected may prove formidable, especially because many galaxies
hosting Compton-thick AGN are also likely to host significant
starburst activity. 

Surveys at submillimeter wavelengths have uncovered a large population 
of luminous, dust-obscured starburst galaxies at 
\hbox{$z\approx 1.5$--3} with star-formation rates (SFRs) of 
the order of \hbox{$\approx 1000$~M$_\odot$~yr$^{-1}$} 
(e.g., Blain et~al. 2002; Chapman et~al. 2003; and 
references therein). Optical spectral 
classification studies of most of these galaxies are difficult due to 
optical faintness, and thus deep \hbox{X-ray} surveys play a critical role in 
determining their AGN content. 
Early comparisons between submillimeter surveys and \hbox{$\approx 20$--150~ks} 
\chandra\ surveys yielded little ($\simlt 10$\%) source overlap, but 
in the exceptionally sensitive CDF-N about 85\% of submillimeter galaxies 
(850~$\mu$m flux densities of 4--12~mJy) with reliable positions
are now detected by \chandra\ (e.g., Barger et~al. 2001c; 
Alexander et~al. 2003a, 2005; Borys et~al. 2004; and references therein). 
The majority of these \hbox{X-ray} detected submillimeter galaxies appear to contain 
moderate-luminosity AGN that are usually obscured (see Figure~9), 
based upon their \hbox{X-ray} spectral shapes and other properties.
Although sample definition and selection effects are complex, the
data suggest an AGN fraction in the submillimeter galaxy population of 
at least 40\% (Alexander et~al. 2005). Thus 
the SMBH in submillimeter galaxies are almost continuously 
growing during the observed phase of intense star formation. 
Even after correcting for the significant amount of \hbox{X-ray} 
absorption present, however, AGN are unlikely to contribute more 
than $\approx 10$--20\% of the bolometric luminosity of typical 
submillimeter galaxies. 


\subsection{X-ray Emission from Distant Starburst and Normal Galaxies}

Although the majority of the \hbox{0.5--10~keV} CXRB has now been resolved, most of 
the extragalactic \hbox{X-ray} sources throughout the Universe have yet to be detected.
These are distant starburst and normal galaxies, where most of the \hbox{X-ray} emission 
arises from \hbox{X-ray} binaries, ultraluminous \hbox{X-ray} sources, 
supernova remnants, starburst-driven outflows, 
and hot gas. Accreting nuclear SMBH (i.e., low-luminosity and heavily
obscured AGN) are certainly also present in some cases (e.g., see Section~2.4), 
although they probably contribute $\simlt 1/3$ of these galaxies'
\hbox{X-ray} emissions at low energies. Some starburst and normal galaxies 
at cosmological distances (\hbox{$z\approx 0.1$--1.5}) are now being detected 
at \hbox{0.5--2~keV} fluxes below 
\hbox{$\approx 5\times 10^{-16}$~erg~cm$^{-2}$~s$^{-1}$}. They constitute 
a rising fraction of the total number of \hbox{X-ray} sources toward fainter 
fluxes (up to \hbox{30--40\%}; see Figure~3), and this trend will continue
until they become the numerically dominant source population at 
\hbox{0.5--2~keV} fluxes of \hbox{$\approx 5\times 10^{-18}$~erg~cm$^{-2}$~s$^{-1}$}
(e.g., Hornschemeier et~al. 2002, 2003; Miyaji \& Griffiths 2002;
Ranalli, Comastri \& Setti 2003; Bauer et~al. 2004). This transition 
from AGN dominance to galaxy dominance of the \hbox{X-ray} number counts is broadly 
analogous to that observed in the radio band below a few millijanskys 
(e.g., Windhorst 2003 and references therein). 

Distant starburst and normal galaxies in \hbox{X-ray} deep fields can be identified
based upon their \hbox{X-ray}, optical, infrared, and radio properties. 
In the \hbox{X-ray} band, these galaxies typically have 
luminosities below $10^{42}$~erg~s$^{-1}$, 
spectra with effective power-law photon 
indices of \hbox{$\Gamma\approx 1.7$--2.2}, and 
X-ray-to-optical flux ratios of $\log (f_{\rm X}/f_{\rm I})<-1$ (see Figure~4). 
Their observed \hbox{X-ray} luminosity function has a lognormal form, as is also
observed for galaxies at infrared and radio wavelengths (Norman et~al. 2004). 
Optically, the distant galaxies detected in X-rays
have relatively high [O~{\sc ii}] $\lambda 3727$ luminosities and morphologies 
consistent with those of field galaxies 
(e.g., Cohen 2003; Hornschemeier et~al. 2003). 
Off-nuclear \hbox{X-ray} sources appear relatively important, as they are 
observed in $\approx 20$\% of these galaxies at \hbox{$z\simlt 0.2$}. 
These have \hbox{X-ray} luminosities of $10^{39}$ to a few times 
$10^{40}$~erg~s$^{-1}$ and are probably luminous black-hole binaries 
or groups of \hbox{X-ray} binaries (e.g., Hornschemeier et~al. 2004). 

The \hbox{X-ray} emission from starburst and normal galaxies can provide
an independent measure of their SFRs that is relatively 
immune to extinction effects (e.g., Bauer et~al. 2002b; Cohen 2003; 
Grimm, Gilfanov \& Sunyaev 2003; Ranalli, Comastri \& Setti 2003; Persic et~al. 2004). 
\hbox{X-ray} derived SFRs agree respectably with SFRs from optical, infrared, and radio 
measurements, at least when the \hbox{X-ray} emission from high-mass \hbox{X-ray} binaries
dominates or can be isolated. Many of the distant galaxies detected in \hbox{X-ray} 
deep fields have remarkably high estimated SFRs of 
$\approx 10$--300~$M_\odot$~yr$^{-1}$. These galaxies are  
members of the strongly evolving, dusty starburst 
population responsible for creating much of the infrared background
(e.g., Alexander et~al. 2002). 

The stellar sources of \hbox{X-ray} emission in galaxies
should show substantial evolution with redshift in response to the factor 
of \hbox{$\approx 10$--100} increase in cosmic SFR out to 
\hbox{$z\approx 1$--3}. The high-mass \hbox{X-ray}-binary population is 
expected to track the cosmic SFR closely, peaking at \hbox{$z\approx 1$--3}, 
whereas low-mass \hbox{X-ray} binaries should track it with a lag of a few Gyr due to 
their longer evolutionary timescales, therefore peaking 
at \hbox{$z\approx 0.5$--1} (e.g., Ghosh \& White 2001). 
Source-stacking analyses have allowed the average \hbox{X-ray} 
properties of $z\approx 0.1$--4 galaxies to be measured,
complementing the individual \hbox{X-ray} detections of galaxies
(e.g., Brandt et~al. 2001ac; Hornschemeier et~al. 2002; Nandra et~al. 2002;
Seibert, Heckman \& Meuer 2002; Georgakakis et~al. 2003; 
Reddy \& Steidel 2004; Lehmer et~al. 2005). 
These analyses indicate that the ratio 
of \hbox{X-ray} to $B$-band luminosity for galaxies rises
from \hbox{$z\approx 0$--1}, such that galaxies at 
$z\approx 1$ are  \hbox{$\approx 2$--5} times as \hbox{X-ray} luminous 
(per unit $B$-band luminosity) as their local counterparts.
Lyman break galaxies at \hbox{$z\approx 2$--4} have typical 
\hbox{X-ray} properties similar to those of the most \hbox{X-ray} 
luminous local starbursts, and \hbox{X-ray} estimates of SFRs
in Lyman break galaxies are in respectable agreement with 
those derived using rest-frame UV data (after appropriate
UV reddening corrections are made). 


\section{Future Prospects for Deep Extragalactic \hbox{X-ray} Surveys}

Follow-up studies for the surveys in Table~1 are ongoing, and both 
\chandra\ and \xmm\ continue to generate torrents of superb data. 
Thus, rapid progress on deep extragalactic \hbox{X-ray} surveys should 
continue over the next decade. Some key outstanding problems requiring 
further work include

\begin{enumerate}

\item
The detailed cosmic history of SMBH accretion. 
Have \hbox{X-ray} surveys found the vast majority of actively accreting
SMBH, or are they missing substantial numbers of Compton-thick and 
other \hbox{X-ray} weak AGN? 
To what degree have the complex \hbox{X-ray} spectra of 
AGN, combined with limited photon statistics, confused current 
estimates of obscuration and luminosity? 
What physical mechanisms are responsible for the observed 
anti-hierarchical growth of SMBH? 

\item
The nature of AGN activity in young, forming galaxies.
How common are moderate-luminosity, typical AGN in the 
$z\approx 2$--10 universe? 
Are these AGN feeding and growing in the same way as local AGN? 
What is the connection between SMBH growth and star formation in 
submillimeter galaxies? 

\item
X-ray measurements of clustering and large-scale structure.  
What are the detailed clustering properties of \hbox{X-ray} selected AGN out 
to high redshift, and is there a dependence of AGN fueling 
upon large-scale environment? How do \hbox{X-ray} groups and clusters 
evolve out to high redshift, and what does this say about 
structure formation? How do \hbox{X-ray}, optical, and radio measures of 
clustering relate? 

\item
The \hbox{X-ray} properties of cosmologically distant starburst and normal 
galaxies. How have the \hbox{X-ray}-source populations in these galaxies
evolved over cosmic history? Is the relationship between \hbox{X-ray}-binary 
production and star formation indeed universal as indicated by the 
correlation between \hbox{X-ray} luminosity and star-formation rate?

\end{enumerate}

\noindent
Table~1 and Figure~1 show that the deep 
\hbox{X-ray} surveys we have reviewed already 
cover a respectable amount of the depth versus solid angle ``discovery space.'' 
However, fully answering the outstanding questions above will likely require 
improvement along both dimensions of this space. 
For example, a longer \hbox{5--10~Ms} \chandra\ observation could 
reach \hbox{0.5--2~keV} flux levels of 
$\approx 5\times 10^{-18}$~erg~cm$^{-2}$~s$^{-1}$ while remaining 
confusion free and nearly photon limited near the field center. 
It would search for distant Compton-thick AGN, 
improve the spectral constraints for and understanding 
of faint \hbox{X-ray}-source populations, and 
detect a few hundred new distant starburst and normal galaxies. 
Such a sensitive \hbox{X-ray} observation will not be possible again 
for \hbox{10--20} years until the launches of missions such as
{\it X-ray Evolving Universe Spectroscopy\/} (\xeus) 
and \genx\ (see Figure~10). 

An equally important approach is to survey more solid angle at 
\hbox{0.5--2~keV} flux levels of 
\hbox{(2--50)$\times 10^{-17}$~erg~cm$^{-2}$~s$^{-1}$}, 
where currently our understanding of the \hbox{X-ray} universe suffers
from limited source statistics and cosmic variance. 
Several surveys underway specifically target this part of 
discovery space, such as 
the Extended \chandra\ Deep Field-South,  
the Extended Groth Strip, and
COSMOS (see Table~1). 
These should substantially improve understanding of
the \hbox{X-ray} luminosity function at high redshift, 
the clustering properties of \hbox{X-ray} selected AGN, and
the evolution of \hbox{X-ray} groups and clusters. 

Finally, a technically challenging but scientifically critical 
goal is to perform deep extragalactic \hbox{X-ray} surveys at higher energies
than those reviewed here. Surveyors of the \hbox{$\approx 10$--100~keV} band
can look forward to observing directly the obscured AGN and other
sources that comprise the bulk of the CXRB. The next decade of research 
in this field should thus be as exciting as the last.


\section*{Acknowledgments}

We thank all of our colleagues for educational interactions on deep 
extragalactic \hbox{X-ray} surveys. Colleagues who specifically helped with 
the preparation of this review include 
DM Alexander, 
FE Bauer, 
H Brunner, 
BD Lehmer, 
V Mainieri,
T Miyaji,  
P Rosati,
DP Schneider, 
AT Steffen, and
C Vignali. 
We acknowledge financial support from NSF CAREER award AST-9983783
and \chandra\ \hbox{X-ray} Center grants (WNB), and DLR grant
50~OR~0207 (GH).  


\begin{figure}
\centerline{\psfig{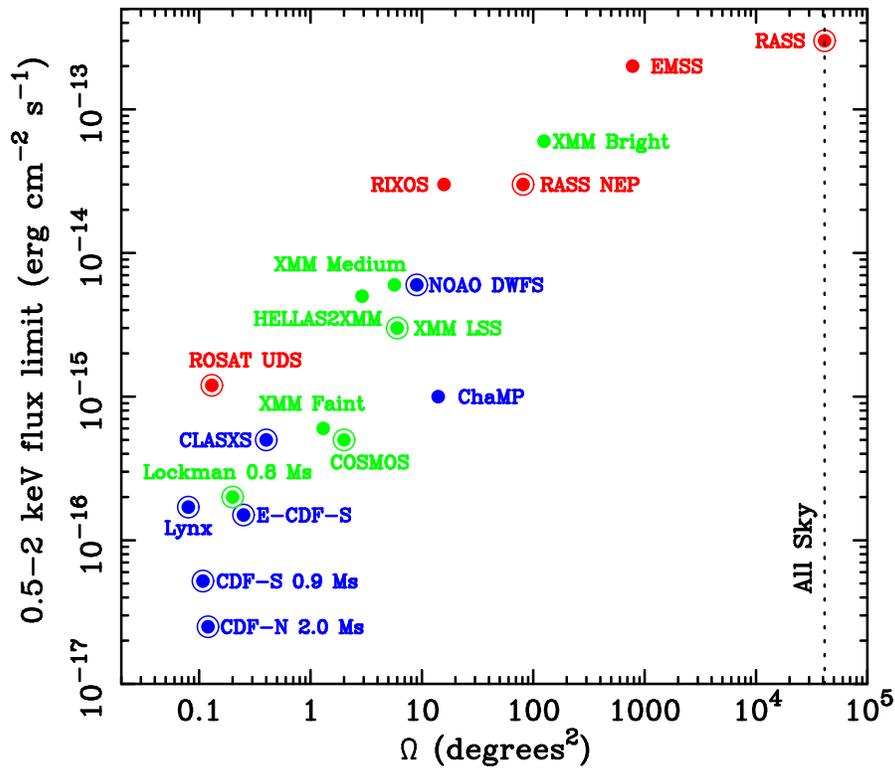}}
\caption{Distributions of some well-known extragalactic surveys 
by \chandra\ (blue), \xmm\ (green), and earlier missions (red) in 
the \hbox{0.5--2~keV} flux limit versus solid angle, $\Omega$, plane. 
Circled dots denote surveys that are contiguous. 
Each of the surveys shown has a range of flux limits across its
solid angle; we have generally shown the most sensitive flux limit. 
The vertical dotted line shows the solid angle of the whole sky.}
\label{fig01}
\end{figure}


\begin{figure}
\centerline{\psfig{figure=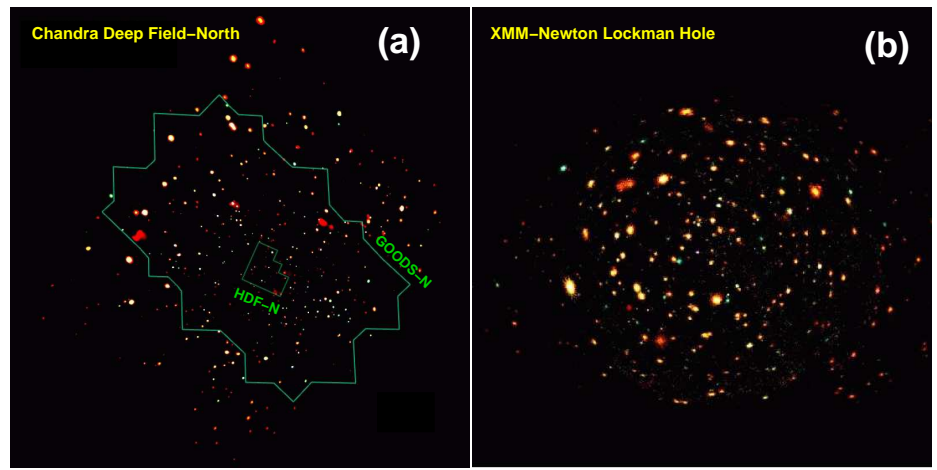,height=4.0 in,angle=-90}}
\caption{(a) The 2.0~Ms \chandra\ Deep Field-North, the deepest \chandra\ observation 
to date. This image has been 
constructed from \hbox{0.5--2~keV} (red), \hbox{2--4~keV} (green), 
and \hbox{4--8~keV} (blue) adaptively smoothed 
images. The two most prominent red diffuse 
patches are galaxy groups/clusters (Bauer et~al. 2002a). 
The regions covered by the \hbox{HDF-N} 
(Ferguson, Dickinson \& Williams 2000) and 
GOODS-N (Giavalisco et~al. 2004) surveys 
with \hst\ are outlined and labeled. This field subtends 
$\approx 448$~arcmin$^2$ ($\approx 60$\% the solid angle of the full Moon), 
and $\approx 580$ sources are detected. 
Adapted from Alexander et~al. (2003b). 
(b) The 770~ks \xmm\ Lockman Hole field, the deepest \xmm\ observation to date. 
This image has been constructed from 0.5--2~keV (red), 2--4.5~keV (green), 
and 4.5--10~keV (blue) images. This field subtends $\approx 1556$~arcmin$^2$, 
and $\approx 550$ sources are detected. 
Adapted from Hasinger (2004).}
\label{fig02}
\end{figure}


\begin{figure}
\centerline{\psfig{figure=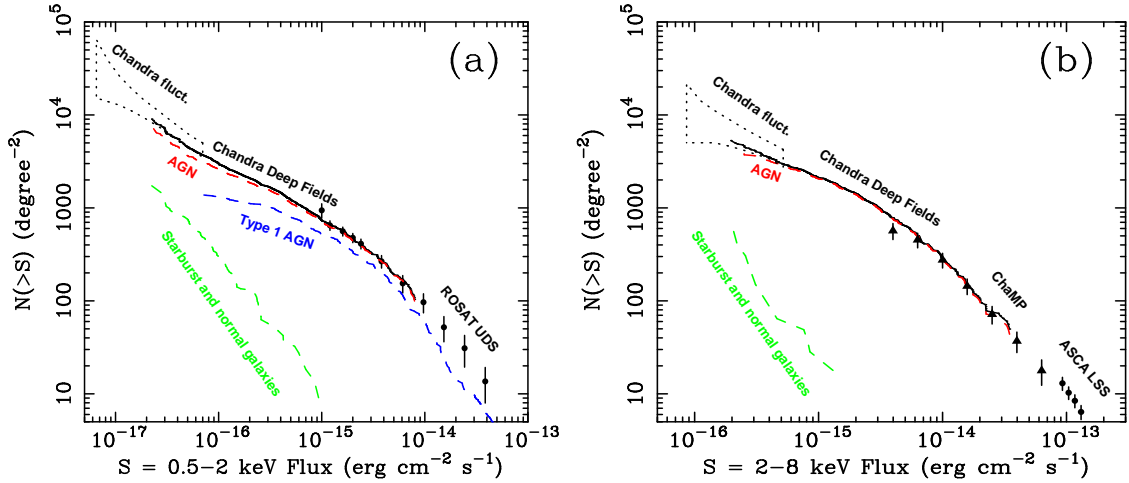,height=2.5 in,angle=0}}
\caption{(a) Number of sources, $N(>S)$, brighter than a given flux, $S$, for the 
0.5--2~keV band. The black circles are from the \rosat\ Lockman Hole 
study of Hasinger et~al. (1998), the solid black curve is from the 
\chandra\ Deep Fields study of Bauer et~al. (2004), and the dotted
black ``fish'' region shows the \chandra\ Deep Field-North fluctuation 
analysis results of Miyaji \& Griffiths (2002). The dashed curves
show number counts for AGN (red), only type~1 AGN (blue), and 
starburst and normal galaxies (green) from Bauer et~al. (2004) and 
Hasinger, Miyaji \& Schmidt (2005). 
(b) $N(>S)$ versus $S$ for the 2--8~keV band. 
The black circles are from the \asca\ Large Sky Survey
study of Ueda et~al. (1999), the black triangles are from the
ChaMP study of Kim et~al. (2004), the solid black curve is from the 
\chandra\ Deep Fields study of Bauer et~al. (2004), and the dotted
black ``fish'' region shows the \chandra\ Deep Field-North fluctuation 
analysis results of Miyaji \& Griffiths (2002). The dashed curves
show number counts for AGN (red) and starburst and normal galaxies
(green) from Bauer et~al. (2004).}
\label{fig03}
\end{figure}


\begin{figure}
\centerline{\psfig{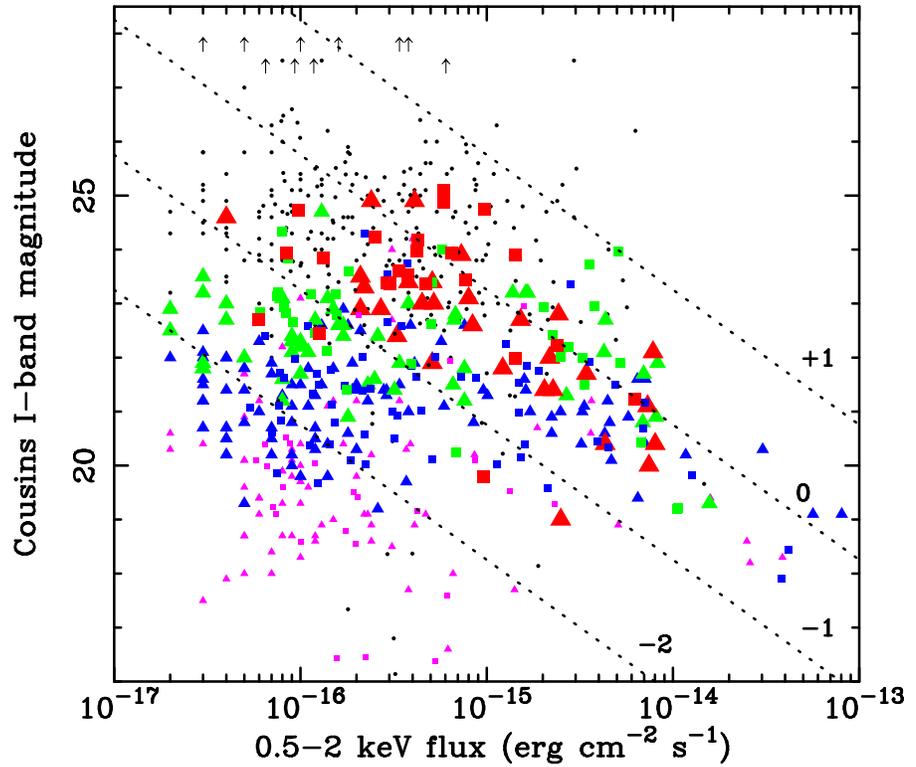}}
\caption{$I$-band magnitude versus \hbox{0.5--2~keV} flux for extragalactic \hbox{X-ray}
sources in the \hbox{CDF-N} (triangles) and \hbox{CDF-S} (squares). 
Sources with redshifts of \hbox{0--0.5}, \hbox{0.5--1}, 
\hbox{1--2}, and \hbox{2--6} are shown as violet, blue, green, and red
symbols, respectively (symbol sizes also increase with redshift). 
Small black dots indicate sources without spectroscopic redshifts. The slanted, dotted
lines indicate constant values of $\log (f_{\rm X}/f_{\rm I})$; the respective
$\log (f_{\rm X}/f_{\rm I})$ values are labeled. Note the broad range of 
optical magnitudes at faint X-ray fluxes and that a large fraction of
the optically faint X-ray sources lack spectroscopic redshifts.}
\label{fig04}
\end{figure}


\begin{figure}
\centerline{\psfig{figure=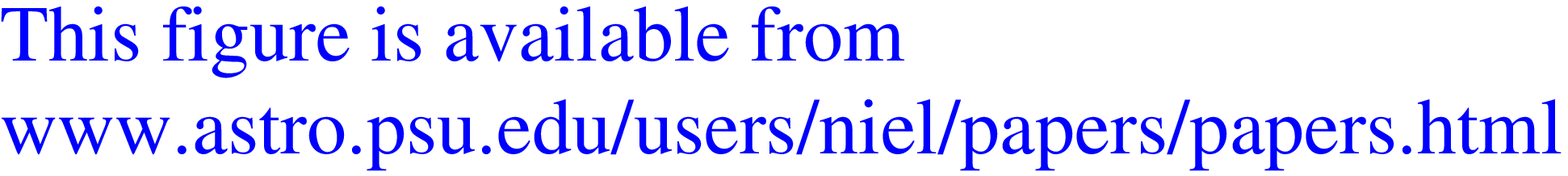,height=4.0 in,angle=-90}}
\caption{\hst\ images of (a) the HDF-N proper 
and (b) the UDF proper with \chandra\ sources circled; 
the circles are much larger than the \chandra\ source positional errors. The numbers 
are source redshifts; redshifts followed by a ``p'' are photometric. Basic source 
type information for many of the sources is also given. The shown HDF-N and UDF 
images subtend $\approx 5.3$~arcmin$^2$ and $\approx 9.5$~arcmin$^2$, respectively.}
\label{fig05}
\end{figure}


\begin{figure}
\centerline{\psfig{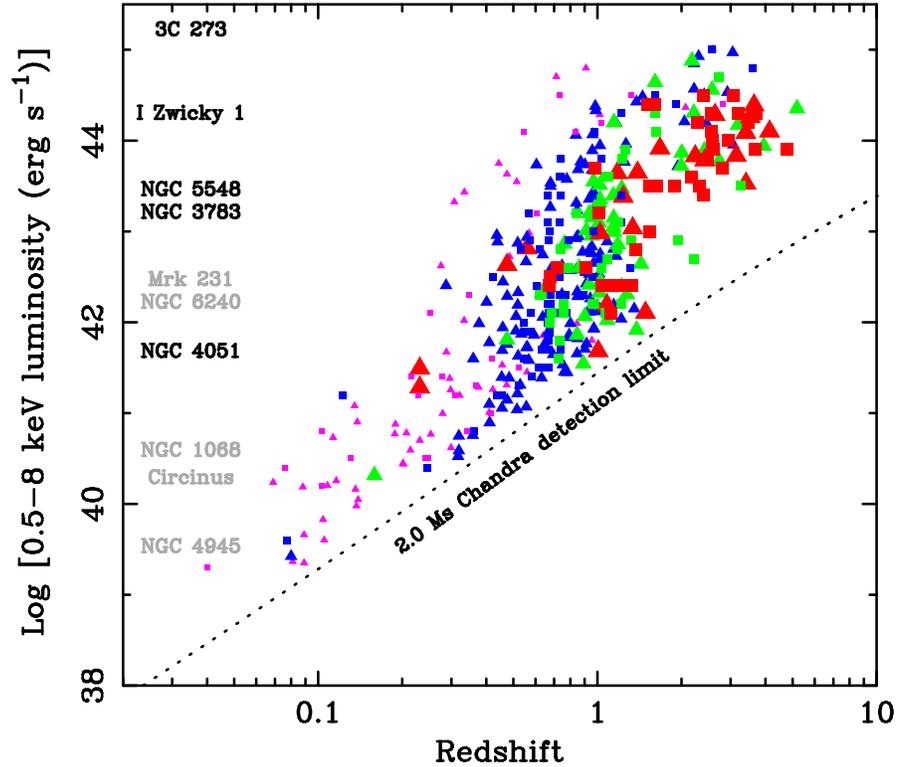}}
\caption{Rest-frame 0.5--8~keV luminosity versus 
redshift for \hbox{CDF-N} (triangles) and \hbox{CDF-S} (squares)
extragalactic sources with spectroscopic redshifts. The dotted curve shows 
the 2.0~Ms sensitivity limit near the center of the \hbox{CDF-N}. Sources 
with \hbox{$I=15$--20}, \hbox{$I=20$--22}, \hbox{$I=22$--23}, and $I>23$ 
are shown as violet, blue, green, and red symbols, respectively (symbol 
sizes also increase with $I$-band magnitude). Note the systematic 
progression of $I$-band magnitudes; the apparent lack of sources with
$z\approx 1.5$--3 and 0.5--8~keV luminosity $\approx 10^{43}$~erg~s$^{-1}$ 
is due to spectroscopic incompleteness at faint $I$-band magnitudes. 
Characteristic \hbox{0.5--8~keV} luminosities for some well-known local AGN are 
shown along the left-hand axis for comparison purposes. Those in 
light gray suffer from Compton-thick absorption and are shown
at the location of their {\it observable\/} \hbox{X-ray} luminosities; their
absorption-corrected luminosities would be \hbox{$\approx 50$--150}
times higher. Note that typical Seyfert~1 galaxies, such as NGC~3783 
and NGC~5548, could have been detected to $z\approx 5$--10 in the \chandra\ 
Deep Fields. Note also that many local Compton-thick AGN would remain 
undetected at $z\simgt 0.2$--1.}
\label{fig06}
\end{figure}


\begin{figure}
\centerline{\psfig{figure=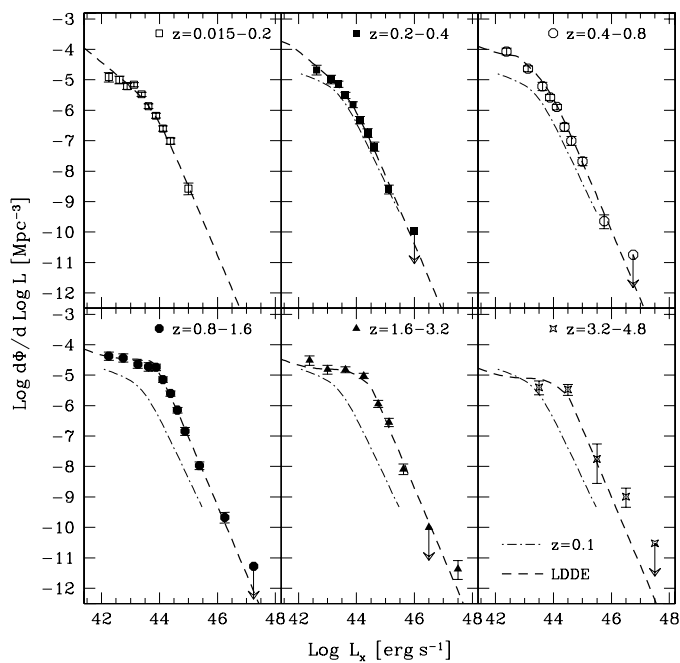,height=3.4 in,angle=0}}
\caption{The \hbox{0.5--2~keV} luminosity function for type~1 AGN in six 
redshift shells. The dashed curves show the best LDDE fit to the data. For 
ease of comparison, the dot-dashed curves in each panel show the best-fit 
model for the \hbox{$z=0.015$--0.2} redshift shell. Adapted from 
Hasinger, Miyaji \& Schmidt (2005).}
\label{fig07}
\end{figure}


\begin{figure}
\centerline{\psfig{figure=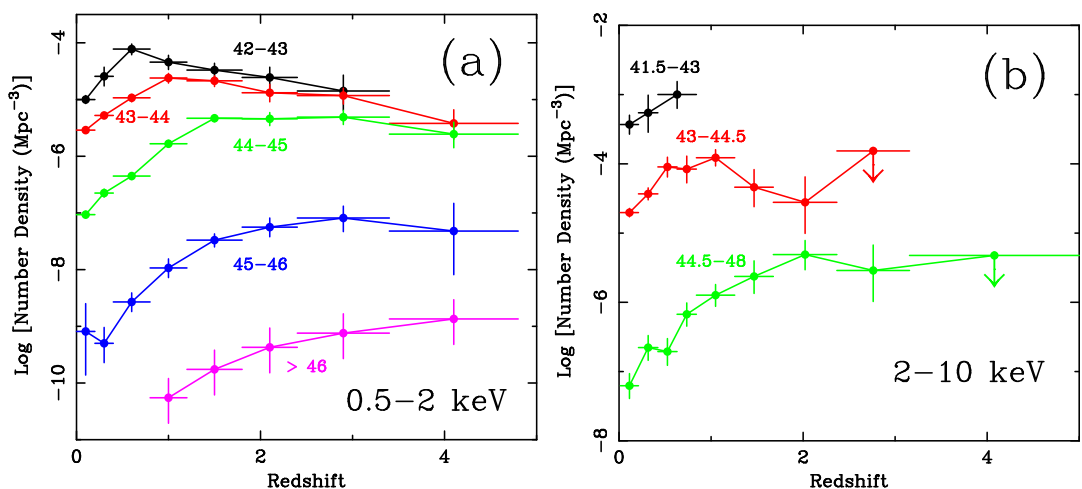,height=2.5 in,angle=0}}
\caption{(a) The comoving space density of AGN selected in the 
\hbox{0.5--2~keV} band as a function of redshift. Results are shown for 
five luminosity ranges; these are labeled with logarithmic luminosity 
values. Adapted from Hasinger, Miyaji \& Schmidt (2005).
(b) The same for AGN selected in the \hbox{2--10~keV} 
band using three luminosity ranges. Adapted from Ueda et~al. (2003).}
\label{fig08}
\end{figure}


\begin{figure}
\centerline{\psfig{figure=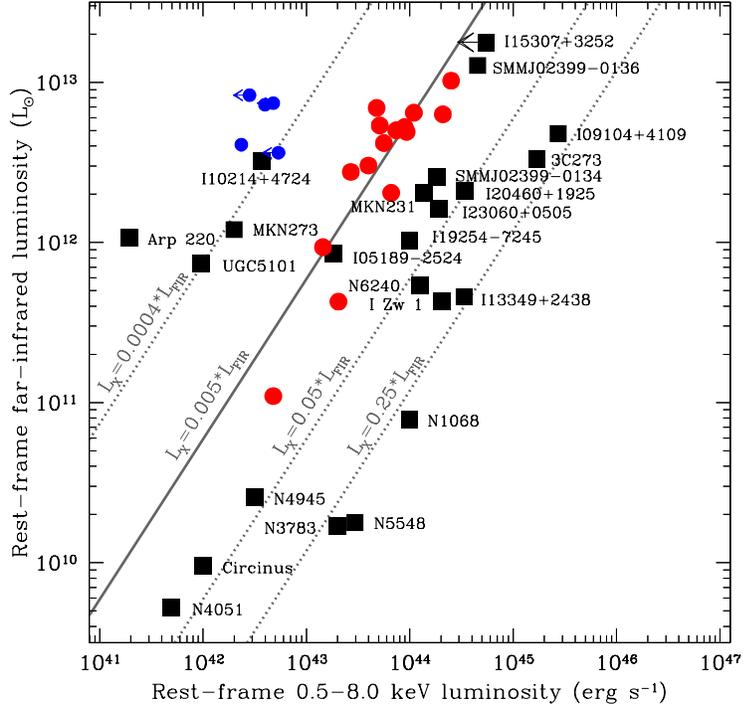,height=4 in,angle=0}}
\caption{Far-infrared luminosity versus absorption-corrected 
X-ray luminosity for submillimeter galaxies in the CDF-N. Red filled circles indicate 
submillimeter galaxies believed to contain AGN, whereas smaller blue filled circles 
indicate submillimeter galaxies that are plausibly pure starbursts. Labeled black 
squares denote well-known AGN-dominated and star-formation
dominated sources from the literature as well as two distant
well-studied submillimeter galaxies containing 
AGN (SMM~02399--0134 and SMM~02399--0136). 
Slanted lines show ratios of constant X-ray to far-infrared luminosity.
This ratio is typically \hbox{$\approx 3$--30\%} for AGN-dominated sources;
many of the CDF-N submillimeter sources appear to contain moderate-luminosity 
AGN, but these AGN are unlikely to dominate the bolometric luminosity. 
Adapted from Alexander et~al. (2005).}
\label{fig09}
\end{figure}


\begin{figure}
\centerline{\psfig{figure=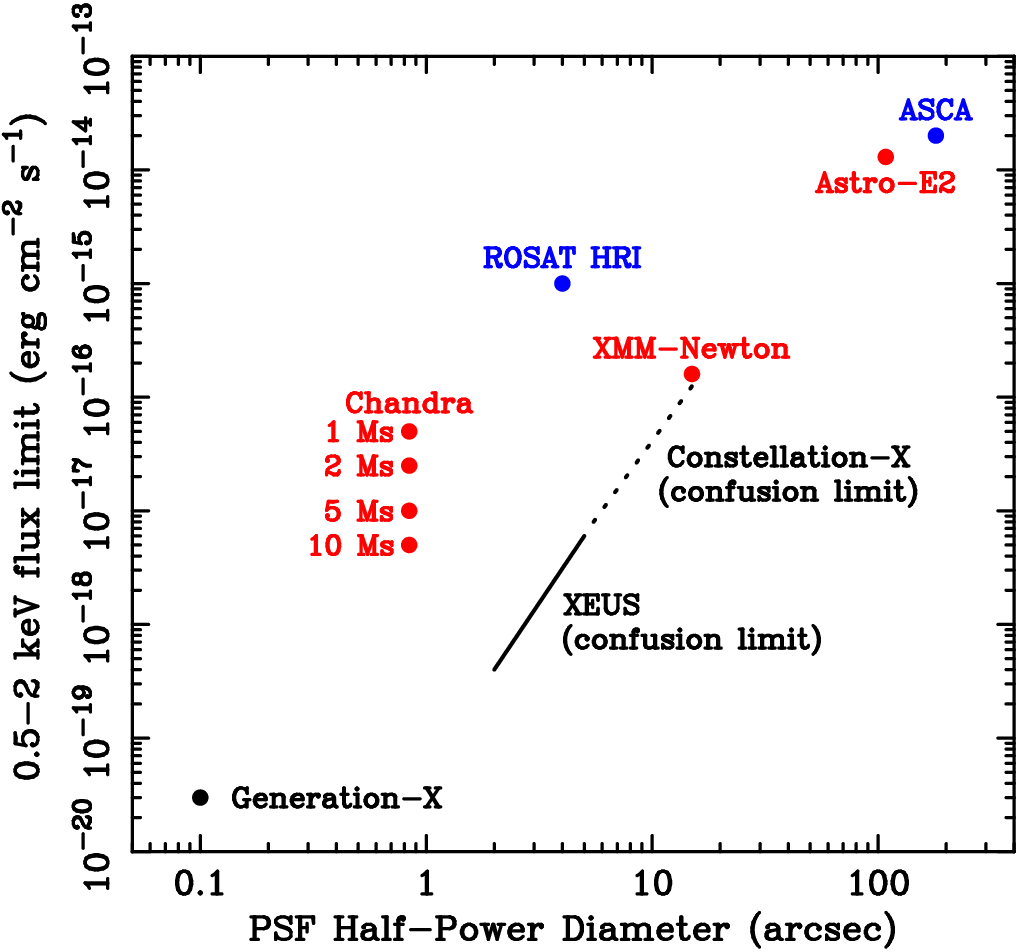,height=3.5 in,angle=-90}}
\caption{\hbox{0.5--2~keV} flux limit versus PSF half-power diameter (HPD) 
for some past (blue), present (red), and future (black) \hbox{X-ray} missions. 
The PSF HPDs for \conx\ and \xeus\ are not precisely determined at 
present, so these missions are shown as diagonal lines (dotted for
\conx\ and solid for \xeus) at the source confusion limits for the range 
of PSF HPDs under consideration. The \genx\ mission is only in a 
preliminary stage of planning at present, and thus its observational 
capabilities have significant uncertainty. 
A $\approx 10$~Ms \chandra\ exposure can achieve sensitivities comparable 
to those planned for \xeus\ and will provide the best available \hbox{X-ray} 
source positions for $\simgt 15$~yr.}
\label{fig10}
\end{figure}



\end{document}